\documentstyle[prl,twocolumn,aps]{revtex} 
\begin{document}
\draft
\title{Importance of core polarization in halo nuclei} 
\author{Attila Cs\'ot\'o}
\address{Theoretical Division, Los Alamos National 
Laboratory, Los Alamos, New Mexico 87545} 
\date{April 16, 1997}

\maketitle

\narrowtext

Recently Kuo, Krpoti\'c, and Tzeng studied the core 
polarization effect in $^6$He and $^{11}$Li \cite{Kuo}. They 
found that the effect is dramatically suppressed in these 
halo nuclei compared to more stable systems. I would like 
to point out that the core polarization, although 
suppressed, still plays an extremely important role in the 
binding mechanism of halo nuclei.

The neutron halo nuclei generally have very small binding
energies relative to breakup thresholds, and the three-body
halos (like $^6$He and $^{11}$Li) usually do not have any 
bound two-body subsystem. These facts indicate that the
most important degrees of freedom are the core$+n(+n)$
relative motion(s). That is why the various cluster models
of these nuclei are so successful \cite{Zhukov}. If one
wants to understand the binding mechanism of halo nuclei
one has to use a model which treats these relative motions
rigorously and, in addition, reproduces the core$+n$ and
$n+n$ scattering observables. It has been known for a long
time that such models underbind the $A=6$ nuclei, including
$^6$He \cite{Lehman}. Moreover, the $^{11}$Li nucleus was 
found to be unbound in such a rigorous three-body model
\cite{Kamimura}, although this fact is less well
established because of the sparse $^9{\rm Li}+n$ data. The
authors of Ref.\ \cite{Kamimura} suggested that if one could
take into account effects where the core is excited through
interacting with one halo neutron and deexcited by
interacting with the other halo neutron, then the missing
binding energy might be recovered. This three-body
excitation-deexcitation mechanism is a physical picture 
of the core polarization effect. 

In Ref.\ \cite{he6} I studied this effect in a six-body,
three-cluster ($\alpha+n+n$) model of $^6$He. In a simple
model, where the alpha particle was not excitable, the
two-neutron separation energy of $^6$He was 0.64 MeV, while
the experimental value is 0.975 MeV. The model reproduced
the experimental $\alpha+n$ and $n+n$ phase shifts, so the
same underbinding problem occured as before \cite{Lehman}.
In a more sophisticated model, the possibility of monopole
$\alpha$-particle excitations were included by allowing
several harmonic oscillator size parameters for the
$\alpha$-particle. It was found that this improvement did 
not have any observable effect on $\alpha+n$ scattering. 
However, the two-neutron separation energy of $^6$He became 
0.74 MeV, 15\% more than in the simpler model. The 
insensitivity of $\alpha+n$ and the sensitivity of 
$\alpha+n+n$ on the inclusion of $\alpha$ excitation into 
the model is a clear, although indirect, indication that 
this 15\% energy gain comes from the core 
excitation-deexcitation mechanism, described above.

In \cite{he6} an even more important core-halo effect was
also found. The $\alpha$-particle can break up and together
with the halo neutrons can form two $^3$H particles. The
inclusion of this effect in the model of \cite{he6} led to
the reproduction of the experimental two-neutron separation
energy of $^6$He in a parameter free model. As the $^9$Li 
core of $^{11}$Li is much softer than the $\alpha$-particle, 
these effects are expected to play a much bigger role in 
$^{11}$Li

In summary, although core polarization is suppressed in
neutron halos, it plays an essential role in the binding 
mechanism of these nuclei.


\begin{references}
\bibitem{Kuo} T.~T.~S. Kuo, F. Krmpoti\'c, and Y. Tzeng,
Phys. Rev. Lett. {\bf 78}, 2708 (1997).
\bibitem{Zhukov} M.~V. Zhukov, B.~V. Danilin, D.~V.
Fedorov, J.~M. Bang, I.~J. Thompson, and J.~S. Vaagen,
Phys. Rep. {\bf 231}, 151 (1993).
\bibitem{Lehman} D.~R. Lehman, Phys. Rev. C {\bf 25}, 3146 
(1982); W.~C. Parke and D.~R. Lehman, {\it ibid.} {\bf 29}, 
2319 (1984).
\bibitem{Kamimura} H. Kameyama, M. Kamimura, and M. Kawai,
in {\it Proceedings of the International Symposium on the 
Structure and Reactions of Unstable Nuclei} (Niigata,
Japan, 1991), edited by K. Ikeda and Y. Suzuki (World
Scientific, Singapore, 1991), p. 203.
\bibitem{he6} A. Cs\'ot\'o, Phys. Rev. C {\bf 48}, 165
(1993).
\end{references}
\end{document}